\documentclass[prb,aps,twocolumn,showpacs]{revtex4}
\usepackage[dvips]{graphicx}
\usepackage{latexsym}
\usepackage{amsmath}
\begin{document}

\title{Quantum spin nematics, dimerization, and deconfined criticality in quasi-one dimensional spin-$1$ magnets}

\author{Tarun Grover$^1$ and T. Senthil$^{1,2}$}
\affiliation{$^1$Center for Condensed Matter Theory, Indian Institute
of Science, Bangalore, India 560012}
\affiliation{$^2$Department of
Physics, Massachusetts Institute of Technology, Cambridge,
Massachusetts 02139}

\date{\today}
\begin{abstract}
We study theoretically the destruction of spin nematic order due to
quantum fluctuations in quasi-one dimensional spin-$1$ magnets. If
the nematic ordering is disordered by condensing disclinations then
quantum Berry phase effects induce dimerization in the resulting
paramagnet. We develop a theory for a Landau-forbidden second order
transition between the spin nematic and dimerized states found in
recent numerical calculations. Numerical tests of the theory are
suggested.

\end{abstract}
\newcommand{\fig}[2]{\includegraphics[width=#1]{#2}}
\newcommand{\be}{\begin{equation}}
\newcommand{\ee}{\end{equation}}
\maketitle

Studies of low dimensional quantum magnetism provide a good
theoretical platform to develop intuition about the physics of
strongly correlated many particle systems. In this paper we study
various theoretical phenomena in spin $S = 1$ quantum magnets with
$SU(2)$ invariant nearest neighbor interactions. Specifically we
focus on spin nematic order in such quantum magnets and its
destruction by quantum fluctuations.

A general Hamiltonian describing such spin $S = 1$ quantum magnets
 takes the form
\begin{equation}
\label{spinham}
 H = \sum_{<ij>} J_{ij} \vec S_i. \vec S_j - K_{ij}
\left(\vec S_i. \vec S_j \right)^2
\end{equation}
In real materials the ratio $K/J$ is probably small; however it has
been proposed that arbitrary values of $K/J$ can be engineered in
ultra-cold atomic Bose gases with spin in optical
lattices\cite{uca}. We will focus exclusively on a rectangular
lattice where the couplings ${J,K}$ on vertical bonds are a factor
of $\lambda$ smaller than those on the horizontal bonds. This model
was studied numerically recently (for $K > 0$) in an interesting paper by Harada
et al\cite{hkt}. In the isotropic limit $\lambda = 1$, they found
that there is a first order phase transition from a collinear Neel
state to a spin nematic state (along the line $J = 0$) with order
parameter
 \be Q_{\alpha\beta} = \left\langle
\frac{S_{\alpha} S_{\beta} + S_{\beta}S_{\alpha}}{2} -
\frac{2}{3} \delta_{\alpha \beta} \right \rangle ~~~\neq 0 \ee even
though there is no ordered moment $<\vec S> = 0$. This spin nematic
state corresponds to the development of a spontaneous hard axis
anisotropy in the ground state. When $\lambda$ is decreased from $1$
to make the lattice rectangular quantum fluctuations are enhanced.
The Neel and spin nematic phases then undergo quantum phase
transitions to quantum paramagnets. Interestingly it is found that
the spin nematic phase gives way to a dimerized quantum paramagnet
where neighboring spin-$1$ moments form strong singlets along every
other bond in the horizontal direction. Further the quantum phase
transition itself appears to be second order in violation of naive
expectations based on Landau theory but similar to the situations
studied in Ref. [\onlinecite{deccp,sandvik}] for other phase transitions in
quantum magnets.

In this paper we provide an understanding of these phenomena. First
we provide general arguments relating the spontaneous dimerization
with one route to killing spin nematic order by quantum
fluctuations. When applied to one dimension our arguments explain
the absence in numerical calculations\cite{nemspch} of the
featureless quantum disordered spin nematic proposed by
Chubukov\cite{andrey} for spin-$1$ chains. Further we show that a
putative direct second order quantum phase transition between the
spin nematic and dimerized phases is described by a continuum field
theory with the action
\begin{eqnarray}
\label{nccp2cont}
 S & = & \int d^3x |\left(\partial_{\mu} -
iA_{\mu}\right) \vec D|^2 +
r|\vec D|^2 + \\ \nonumber
& & u \left(|\vec D|^2\right)^2 - v (\vec D)^2(\vec D^*)^2 +
\frac{1}{e^2}\left(\epsilon_{\mu\nu\lambda} \partial_{\nu}
A_{\lambda} \right)^2
\end{eqnarray}
Here $\vec D$ is a complex three component vector and $A_{\mu}$ is a
non-compact $U(1)$ gauge field. The nematic phase occurs when $\vec
D$ condenses while $\vec D$ is gapped in the dimerized paramagnet.
This theory is an anisotropic version of the non-compact $CP^2$
model ($NCCP^2$). The two component version - the anisotropic
$NCCP^1$ model -  describes Neel-VBS transitions of easy plane
spin-$1/2$ magnets on the square lattice\cite{deccp}. A second order
nematic-dimer transition on the rectangular lattice is possible if
{\em doubled} instantons in the gauge field $A_{\mu}$ are irrelevant
at the critical fixed point of this theory. These instantons are
relevant at the paramagnetic fixed point of Eqn. \ref{nccp2cont}. This
leads to confinement of the $\vec D$ fields and 
to dimer order. Indeed the dimer order parameter is simply the single instanton operator. 
A direct second order nematic-dimer transition
is thus accompanied by the dangerous irrelevance of doubled
instantons and the associated two diverging length/time scales.

We will provide two different arguments to justify our results.
First we address the quantum disordering of the nematic based on
general effective field theory considerations that focus on the
properties of topological defects of the nematic order parameter.
Second we provide a more microscopic argument based on an exact
slave `triplon' representation\cite{slvtrp} of the $S = 1$
operators.

The order parameter manifold for the spin nematic state may be taken
to be the possible orientations of the spontaneous hard axis
$\hat{d}$ (the ``director") and thus is $S^2/Z_2$. The $Z_2$ simply
reflects the fact that $\hat{d}$ and $-\hat{d}$ are the same state.
The spin nematic state  allows for $Z_2$ point vortex defects
(``disclinations") in two space dimensions. The director $\hat{d}$
winds by $\pi$ on encircling such a disclination. As discussed by
Lammert et al\cite{lammert} for classical nematics the fate of the
disclinations crucially determines the nature of the ``isotropic"
phase obtained when the nematic is disordered by fluctuations. If
the transition out of the nematic occurs without condensing the
disclinations then a novel topologically ordered phase - interpreted
in the present context as a quantum spin liquid - obtains. However
the nematic may also be disordered in a more conventional way by
condensing the disclinations. In the present context we argue that
non-trivial quantum Berry phases associated with the disclinations
lead to broken translational symmetry in this quantum paramagnet.
Furthermore this transition may be second order as described below
(unlike the classical nematic-isotropic transition).

Following Lammert et al\cite{lammert} we consider the quantum phase
transition out of the nematic using an effective model in terms of
the director $\hat{d}$. The $\hat{d}, -\hat{d}$ identification
requires that the $\hat{d}$ vector is coupled to a $Z_2$ gauge
field. Thus we consider the following action on a three dimensional
spacetime lattice:
\begin{eqnarray}
\label{lgm}
S & = & S_d + S_B \\
S_d & = & -\sum_{<r, r+\mu>} t^d_{\mu}\sigma_{\mu}(r) \hat{d}_r.
\hat{d}_{r+\mu}
\end{eqnarray}
Here $r$ represent the sites of a cubic spacetime lattice, $\mu =
(x, y, \tau)$, $\sigma_{\mu}(r) = \pm 1$ is a $Z_2$ gauge field on
the link between $r$ and $r +\mu$. The term $S_B$ is the Berry phase
to be elaborated below.

The Berry phases arise from the nontrivial quantum dynamics of the
$\hat{d}$ vector and can be understood very simply by considering a
single quantum spin $S = 1$ with a time varying hard axis
$\hat{d}(\tau)$ that represents the fluctuating local director
field:
\begin{equation}
{\cal H} = \left(\hat{d}(\tau).\vec S \right)^2
\end{equation}
For a {\em time independent} $\hat{d}$, the ground state is simply
the state where the projection $\vec S.\hat{d} = 0$. The Berry phase
is obtained by considering a slow time varying closed path of
$\hat{d}$ in the adiabatic approximation. There are two kinds of
such closed paths that are topologically distinct. First there are
paths for which $\hat{d}$ returns to itself. For such paths it is
easy to see that the Berry phase factor is $1$. Then there are
closed paths where $\hat{d}$ returns to $-\hat{d}$. In the adiabatic
approximation with $S = 1$ it is easy to see that the wavefunction
acquires a phase of $\pi$ for such a path. Thus there is a Berry
phase of $-1$ for closed paths where $\hat{d}$ returns to
$-\hat{d}$.

The phase factor of $-1$ for nontrivial closed time evolutions of
$\hat{d}$ at a spatial site may be naturally incorporated into the
effective lattice model of Eqn. \ref{lgm} above. First we note that
a closed loop in time where $\hat{d}$ winds by $\pi$ corresponds to
a configuration with $Z_2$ gauge flux $-1$ through the loop. The
Berry phase is thus simply
\begin{equation}
e^{-S_B} = \prod_{r} \sigma_{\tau}(r)
\end{equation}
At each space point the product over the time-like bonds measures
the flux of the $Z_2$ gauge field through the closed time loop at
that point. Precisely this Berry phase factor arises in $Z_2$ gauge
theoretic formulations of a number of different strong correlation
problems\cite{z2long}, and the theory is known as the odd $Z_2$
gauge theory. Thus an appropriate effective model for disordering
the $S = 1$ spin nematic state is a theory of $\hat{d}$ coupled to
an odd $Z_2$ gauge theory.

The spin nematic ordered phase corresponds to a condensate of
$\hat{d}$. In this phase the $Z_2$ disclinations are simply
associated with $Z_2$ flux configurations of the gauge field. Thus
the Berry phase term associated with the gauge field directly
affects the dynamics of the disclinations. Disordered phases where
$\hat{d}$ has short ranged correlations may be discussed by
integrating out the $\hat{d}$ field. The result is pure odd $Z_2$
gauge theory on a spatial lattice with rectangular symmetry. This
theory is well understood. It is conveniently analysed by a duality
transformation to a stacked fully frustrated Ising model\footnote{The frustration in the Ising model is directly due to the Berry phase in the gauge theory} followed by
a soft spin Landau-Ginzburg analysis\cite{fftfim}. This leads to a
mapping to an $XY$ model with four-fold anisotropy:
\begin{equation}
\label{vrtxy}
 S_v = -t_v \sum_{<RR'>} cos(\phi_R - \phi_{R'}) - \kappa
\sum_R cos(4\phi_R)
\end{equation}
Here $R, R'$ are sites of the dual cubic lattice. The real and
imaginary parts of the field $e^{i\phi_R}$ correspond to Fourier
components of the $Z_2$ vortex near two different wavevectors at
which the quadratic part of the Landau-Ginzburg action has minima.
The anisotropy is four-fold on the rectangular spatial lattice as
opposed to the $8$-fold anisotropy that obtains with square
symmetry\cite{fftfim}. There is a disordered phase where the $Z_2$
vortex has short-ranged correlations: this corresponds to the
topologically ordered quantum spin liquid in the original spin
model. In addition there are ordered phases associated with
condensation of the $e^{i\phi_R}$. These phases break translation
symmetry. For the rectangular lattice of interest the natural
symmetry breaking pattern is dimerization along the chain direction.

We thus see that Berry phases associated with the quantum dynamics
of the director $\hat{d}$ lead to dimerization when the nematic
order is disordered by condensing the $Z_2$ disclinations. This
analysis can be easily repeated in one spatial dimension. Then the
$Z_2$ disclinations are point defects in spacetime. These are
described by the odd $Z_2$ gauge theory in $1+1$ dimensions which is
always confined and which has a dimerized ground state\cite{msf}. In
particular this argument shows that for $S = 1$ chains a featureless
disordered spin nematic state will not exist.

Returning to two dimensions we may now write down a field theory for
the nematic-dimer transition. The Berry phases on the disclinations
are encapsulated in Eqn. \ref{vrtxy}. We now need to couple these
back to the $\hat{d}$ vector. The main interaction between $\hat{d}$
and $e^{i\phi_R}$ is the long ranged statistical one: on going
around a particle created by $e^{i\phi}$ the vector $\hat{d}$
acquires a minus sign.


We proceed by first ignoring the $\kappa$ term in Eqn. \ref{vrtxy}
and using a Villain representation to let
\begin{equation}
S_v = \sum_{<RR'>} U j_{RR'}^2
\end{equation}
The $j_{RR'}$ are integer valued currents of the $e^{i\phi}$ that
satisfy the conditions
\begin{eqnarray}
\label{cons}
\vec \nabla. \vec j & = & 0 \\
\label{mutstat}
 (-1)^j  & = & \prod_P \sigma
\end{eqnarray}
The first equation expresses current conservation (which holds at
$\kappa = 0$). In the second the symbol $\prod_P$ refers to a
product over the four bonds of the direct lattice pierced by
$<RR'>$.  This term ensures that an $e^{i\phi}$ particle acts as
$\pi$ flux for the $\hat{d}$ field. Solving Eqn. \ref{cons} by $\vec
j = \vec \nabla \times \vec A$ (with $A_{\mu}$ integer) we get
\begin{equation}
(-1)^{\vec \nabla \times \vec A} = \prod_P \sigma
\end{equation}
Now write $A = 2a + s$ with $a$ an integer and $s = 0, 1$ so that
\begin{equation}
\prod_P (-1)^s = \prod_P \sigma
\end{equation}
which can be solved by choosing
\begin{equation}
(-1)^s = 1-2s~~= \sigma
\end{equation}
The integer constraint on $A$ may be implemented softly by including
a term
\begin{equation}
-t_{\theta}cos(2\pi a)  =   -t_{\theta} \sigma_{rr'} cos(\pi
A_{rr'})
\end{equation}
We now separate out the longitudinal part of $A$ by letting
\begin{equation}
\vec A \rightarrow \vec A + \frac{1}{\pi}\vec \nabla \theta
\end{equation}
After a further rescaling $A \rightarrow \frac{A}{\pi}$ we finally
get the action
\begin{eqnarray}
S & = & S_d + S_{\theta} + S_A \\
S_{\theta} & = & -t_{\theta} \sum_{<rr'>} \sigma_{rr'} cos(\theta_r
-
\theta_{r'} + A_{rr'}) \\
S_A & = & U \sum_P \left( \vec \nabla \times \vec A \right)^2
\end{eqnarray}
with $S_d$ given in eqn. \ref{lgm}. The sum over $\sigma$ can now be
performed. The universal prperties are correctly captured in a model that keeps the 
lowest order cross term between $t^d$
and $t_{\theta}$. We therefore get
\begin{equation}
\label{dthetamod}
 S  =  -\sum_{<rr'>} t_{\mu}cos(\theta_r - \theta_{r'} +
A_{rr'})\hat{d}_r. \hat{d}_{r'} + U\sum_P (\vec \nabla \times \vec
A)^2
\end{equation}
with $t_{\mu} \sim t^d_{\mu} t_{\theta}$.  It is instructive to
introduce the complex vector $\vec D_r = e^{i\theta_r} \hat{d}_r$
that satisfies
\begin{eqnarray}
|\vec D|^2 & = & 1 \\
\vec D \times \vec D^* & = & 0
\end{eqnarray}
The second condition may be imposed softly by including a term
\begin{equation}
v|\vec D \times \vec D^*|^2  =  -v \left((\vec D)^2(\vec D^*)^2 -1
\right)
\end{equation}
with $v > 0$. Thus we arrive at the model
\begin{eqnarray}
\label{lattnccp2}
S & = & S_D + S_A \\
S_D & = & -t\sum_{<rr'>}  e^{iA_{rr'}} \vec D^*_r. \vec D_{r'} + c.c
-v (\vec D)^2(\vec D^*)^2
\end{eqnarray}
with $\vec D$ satisfying $|\vec D|^2 = 1$. At $v = 0$ this is the
lattice $CP^2$ model with a global $SU(3)$ symmetry associated with
rotations of the $\vec D$ field. The $v$ term breaks the $SU(3)$
symmetry down to $SO(3)$. Eqn. \ref{nccp2cont} is precisely a
soft-spin continuum version of the lattice action above.

We now consider the role of the four fold anisotropy on the
disclination field $e^{i\phi}$ (the $\kappa$ term in Eqn
\ref{vrtxy}). Without this term the number conjugate to $\phi$ is
conserved. In the dual description this translates into conservation
of the magnetic flux of the $U(1)$ gauge field $\vec A$. Thus at
$\kappa = 0$ the gauge field is noncompact. The $\kappa$ term
however destroys this conservation law - indeed four disclinations
can be created or destroyed together. In the effective model of Eqn.
\ref{nccp2cont}, a disclination in $\vec D$ corresponds to a
configuration where the gauge flux is equal to $\pi$. Thus the
$\kappa$ term may be interpreted as a doubled `instanton' operator
that changes the gauge flux by $4\pi$.

The nematic order parameter is simply related to the $\vec D$
fields: \be Q_{\alpha\beta} = \left(\frac{D^*_\alpha D_\beta +
c.c}{2} - \frac{\delta_{\alpha\beta}}{3}\right) \ee Thus when $\vec
D$ condenses nematic order develops. The paramagnetic phase occurs
when $\vec D$ is gapped. In the absence of instantons the low energy
theory of this phase has a free propagating massless photon.
Instantons however gap out the photon and confine the $\vec D$ fields. The dimer order parameter $e^{i\phi}$ is the single instanton operator
and gets pinned in this phase. A
direct second order transition between the nematic and dimerized
states can thus occur if  doubled instantons are irrelevant at the
critical fixed point of the anisotropic $NCCP^2$ action  associated
with the condensation of $\vec D$.

A different more microscopic argument can also be used to justify
Eqn. \ref{nccp2cont} and provides further insight. Consider the
following exact representation\cite{slvtrp} of a spin-$1$ operator
at a site $i$ in terms of a `slave' triplon operator $\vec w_i$: \be
\vec S_i = -i \vec w^{\dagger}_i \times \vec w_i \ee together with
the constraint $\vec w^{\dagger}_i.\vec w_i = 1$. The $\vec w_i$
satisfy usual boson commutation relations.  The nematic order
parameter is readily seen to simply be
\begin{equation}
Q_{\alpha\beta} = \left\langle
\frac{\delta_{\alpha\beta}}{3}-\frac{w^\dagger_\alpha w_\beta +
c.c}{2}\right\rangle
\end{equation}
As with other slave particles, this representation leads to a $U(1)$
gauge redundancy associated with letting \be \vec w_i \rightarrow
e^{i\alpha_i} \vec w_i \ee at each lattice site. It is convenient to
first consider the special point $J_{ij} = 0$ where the Hamiltonian
in Eqn. \ref{spinham} is known to have extra $SU(3)$ symmetry. Then
$H$ may be rewritten (upto an overall additive constant)
\begin{equation}
H = - \sum_{<ij>} K_{ij}\left(\vec w^{\dagger}_i. \vec w^{\dagger}_j
\right) \left(\vec w_i. \vec w_j \right)
\end{equation}
This is invariant under a global multiplication of $\vec w_i$ by an
$SU(3)$ matrix $U$ on one sublattice and by $U^*$ on the other. Such
magnets were studied in detail in Ref. \onlinecite{ReSaSuN} and we
can take over many of their results. A standard mean field
approximation with $<\vec w_i. \vec w_j> \neq 0$ yields a
paramagnetic phase with gapped $\vec w$ particles in the $d = 1$
limit while in two dimensions the $\vec w$ condense thereby breaking
the $SU(3)$ symmetry. The theory of fluctuations beyond mean field
includes a compact $U(1)$ gauge field. In the paramagnetic phase
instanton fluctuations of this gauge field confine the $\vec w$
particles and their Berry phases lead to dimerization on the
rectangular lattice. The results of Ref. \onlinecite{ReSaSuN} now
imply that the transition associated with $\vec w$ condensation is
described by an $NCCP^2$ model with doubled instantons, {\em i.e} it
is precisely of the form of Eqn. \ref{nccp2cont} but with $v = 0$.
The triplon $\vec w_i$ on the A sublattice $\sim \vec D$ while on
the other sublattice $\vec w_i \sim \vec D^*$. Thus we see that the
Neel vector $\vec N$ is simply related to $\vec D$ through
\begin{equation}
\label{neel}
\vec N \sim -i \vec D^* \times \vec D
\end{equation}
For the $SU(3)$ symmetric Hamiltonian all eight components of the
tensor $D^*_\alpha D_\beta - |D|^2 \delta_{\alpha \beta}/3$ have the
same correlators. The symmetric part of this tensor is the nematic
order parameter and the antisymmetric part is the Neel vector.

If now a small $J < 0$ is turned on the $SU(3)$ symmetry is
explicitly broken down to $SO(3)$. This sign of $J$ disfavors Neel
ordering so that nematic ordering wins in the two dimensional limit.
The $NCCP^2$ field theory of the transition to the dimer state must
then be supplemented with an anisotropy term $v|\vec N|^2$
with $v
> 0$ which due to Eqn. \ref{neel} is precisely the anisotropy term
of Eqn. \ref{nccp2cont}.

What may we say about the $NCCP^2$ field theory and the fate of
doubled instantons? First as the critical theory is relativistic the
dynamical critical exponent $z = 1$. Next we note that the instanton
scaling dimension is expected to be bigger for $NCCP^2$ as compared
to $NCCP^1$. In the isotropic case existing estimates\cite{sandvik}
give $0.63$ for the single instanton scaling dimension. In a naive
RPA treatment of the gauge fluctuations the instanton scaling
dimension scales like $m^2 N$ where $m$ is the instanton charge and
$N$ is the number of boson components. Thus within this
approximation we estimate the doubled instanton scaling dimension in
$NCCP^2$ as $ \frac{3}{2}(2)^2(.63) \approx 3.78$. This admittedly
crude estimate nevertheless suggests that doubled instantons may be
irrelevant for $NCCP^2$.

The possible irrelevance of the doubled instantons has dramatic
consequences for the phenomena at the nematic-dimer transition. It
implies that the critical fixed point has enlarged $U(1)$ symmetry
associated with conservation of the gauge flux exactly like in Ref.
\onlinecite{deccp}. This enlarged symmetry implies that the
$(\pi,0)$ columnar dimer order parameter may be rotated into the
$(0, \pi)$ columnar dimer or into plaquette order parameters at
$(0,\pi)$, $(\pi,0)$. Thus right at the critical point all these
different VBS orders will have the same power law correlations. It
will be an interesting check of the theory of this paper to look for
this in future numerical calculations.

In summary we have studied the destruction of spin nematic order by
quantum fluctuations in quasi-one dimensional spin-$1$ magnets. We
showed that Berry phases associated with disclinations lead to
dimerization if the nematic is disordered by their condensation. We
presented a continuum field theory for a Landau-forbidden second
order transition between nematic and dimerized phases that
generalizes earlier work on other transitions. Future numerical work
or cold atoms experiments may be able to explore the physics
described in this paper.

TS gratefully acknowledges support from the DAE-SRC Outstanding
Investigator Program in India.

\bibliography{}
\end{document}